\documentclass[12pt,preprint]{aastex}

\begin{document}

\title{Global X-ray properties of the Orion Nebula region}

\author{Eric D.\ Feigelson\altaffilmark{1}, Konstantin
Getman\altaffilmark{1}, Leisa Townsley\altaffilmark{1}, Gordon
Garmire\altaffilmark{1}, Thomas Preibisch\altaffilmark{2}, Nicolas
Grosso\altaffilmark{3}, Thierry Montmerle\altaffilmark{3}, Augustus
Muench\altaffilmark{4}, and Mark McCaughrean\altaffilmark{5,6}}

 \altaffiltext{1}{Department of Astronomy and Astrophysics,
Pennsylvania State University, 525 Davey Laboratory, University
Park, PA 16802}
 \altaffiltext{2}{Max-Planck-Institut f\"ur
Radioastronomie, Auf dem H\"ugel 69, D-53121 Bonn, Germany}
 \altaffiltext{3}{Laboratoire d'Astrophysique de
Grenoble, Universit{\'e} Joseph-Fourier, F-38041 Grenoble cedex 9,
France}
 \altaffiltext{4}{Harvard-Smithsonian Center for Astrophysics,
60 Garden Street, Cambridge MA 02138}
 \altaffiltext{5}{University of Exeter, School of Physics, Stocker
Road, Exeter EX4 4QL, Devon, UK}
 \altaffiltext{6}{Astrophysikalisches Institut Potsdam, An der 
Sternwarte 16, D-14482 Potsdam, Germany}

\slugcomment{Resubmitted to Astrophys. J. Suppl.}

\begin{abstract}

Based on the $Chandra$ Orion Ultradeep Project (COUP) observation, we
establish the global X-ray properties of the stellar population
associated with the Orion Nebula.  Three components contribute
roughly equally to the integrated COUP luminosity in the hard ($2-8$
keV) X-ray band:  several OB stars, 822 lightly obscured cool stars
in the Orion Nebula Cluster (ONC), and 559 heavily obscured stars.
ONC stars $0.5-2$~pc from the center show a spatial asymmetry
consistent with violent relaxation in the stellar dynamics.  The
obscured COUP sources concentrate around both OMC-1 molecular cores;
these small-scale structures indicate ages $t \la 0.1$ Myr.  The
X-ray luminosity function (XLF) of the lightly obscured sample is
roughly lognormal in shape.  The obscured population is deficient in
lower-luminosity stars, perhaps due to localized circumstellar
material.  Mass-stratified XLFs show that one-third of the Orion
Nebula region hard-band emission is produced by the bright O6 star
$\theta^1$ Ori C and half is produced by lower mass pre-main sequence
stars with masses $0.3<M<3$ M$_\odot$.  Very low mass stars
contribute little to the cluster X-ray emission.

Using the hard band emission, we show that young stellar clusters
like the ONC can be readily detected and resolved with $Chandra$
across the Galactic disk, even in the presence of heavy obscuration.
The Orion Nebula sample is a valuable template for studies of distant
clusters.  For example, the peak of the XLF shape can serve as a
standard candle for a new distance measure to distant young stellar
clusters, and the presence of a neon emission line complex around 1
keV can serve as a diagnostic for young stars.

\end{abstract}

\keywords{stars: formation -- stars: pre-main sequence -- ISM:
individual (Orion Nebula) -- open clusters and associations:
general -- open clusters and associations: individual (Orion
Nebula Cluster) -- X-rays: stars}

\section{Introduction \label{intro.sec}}

Star formation regions are generally studied at long wavelengths.
Molecular clouds cool by far infrared emission of dust and 
rotational lines such as the $J=2 \rightarrow 1$ line of CO.  Young 
stellar clusters (YSCs) are investigated directly at optical and 
near-infrared wavelengths if they are relatively unobscured, and at 
mid- to far-infrared if they are deeply embedded so that their 
starlight is reprocessed by heated dust.  Gas in H{\sc II} regions 
heated by massive stars can be studied at radio through ultraviolet 
wavelengths, but give limited information about the full stellar 
population.

It may seem at first that X-ray observations would make only very
limited contributions to star formation and YSC studies.  Lower mass
pre-main sequence (PMS) stars typically emit only $10^{-4}$ of their
bolometric radiation in the X-ray band \citep{Feigelson99, Favata03,
Guedel04}, and the ratio is even smaller for OB stars where $\log
L_x/L_{bol} \simeq - 7$.  Yet, modern X-ray telescopes such as the
$Chandra$ $X-ray$ $Observatory$ and the $XMM-Newton$ satellite are
surprisingly sensitive.  For a $Chandra$ exposure of 100 ks, a
single PMS star with $L_{bol} = 1$ L$_\odot$ can be detected with 10
photons out to 2 kpc and an O star with $L_{bol} = 1 \times 10^4$
L$_\odot$ can be detected out to 6 kpc, assuming little absorption.
A compact massive YSC with an Initial Mass Function (IMF) extending
from early-O stars through brown dwarfs will typically emit $\simeq
1- 10$ L$_\odot$ in the hard X-ray band which can be detected
through most of the Galactic disk.

Obscuration by the natal molecular cloud and, for distant star
forming regions, intervening spiral arms, presents the major
difficulty at all wavelengths for studying star formation on a
Galactic scale.  X-ray astronomy can help because a significant
fraction of the X-ray photons from both low and high mass stars are
emitted at energies above 3 keV which can penetrate through nearly
all lines-of-sight in the Galaxy and deep into molecular clouds. For 
example, in the $Chandra$ Orion Ultradeep Project (COUP) discussed 
here, the protostar IRS~2 in the OMC-1 South molecular core was 
detected with 30 cts/(100~ks) with an X-ray-derived line-of-sight 
column density $\log N_H \simeq 23.9$ cm$^{-2}$ equivalent to $A_V 
\sim 500$ mag \citep{Grosso05}.

X-ray studies of young stars in star formation regions provide a
variety of insights that complement those obtained at longer
wavelengths.  These include:  \begin{enumerate}

\item X-ray images give a census of members of YSCs that is
largely independent of the presence or absence of a dusty 
circumstellar disk, thereby overcoming the bias of memberships 
derived from infrared-excess criteria \citep{Feigelson99}. 
Sufficiently deep X-ray images can peer into the lower mass 
population in YSCs where only the OB stars have been studied.  Due 
to a statistical correlation between X-ray luminosity and mass 
\citep{Preibisch05a}, the X-ray luminosity function (XLF) might be 
useful to test uniformity of the stellar Initial Mass Function (IMF) 
in different clusters \citep{Feigelson05b}.  Furthermore, the 
monotonic decrease of $\sim 10^2$ in X-ray emission between young 
stars with ages $5 < \log t < 10$ yr -- from the younger Orion stars 
into the main sequence  \citep{Preibisch05b} and continuing to the 
oldest disk stars \citep{Feigelson04} -- can greatly assist the 
removal of contaminants from a YSC membership census.

\item The shape of the stellar XLF in YSCs appears to be universal, 
and can probably be used as a standard candle for distance 
determinations \citep{Feigelson05b}.  The distances to YSCs which 
are too poor or obscured to exhibit a clear OB main sequence are 
often very uncertain.

\item High-resolution X-ray imagery can detect bolometrically
faint stars in the close vicinity of bolometrically bright stars,
because the contrast in X-ray brightnesses is often small
\citep{Stelzer05}. For example, a short $Chandra$ observation
discovered four companions within 10\arcsec\/ of a $V \simeq 7$ PMS
star \citep{Feigelson03}.

\item X-ray studies of YSCs with well-characterized individual
stars permit study of magnetic activity of PMS stars as a function
of mass, rotation and age \citep{Preibisch05a}.  X-ray flares give
particularly important insight into high energy inputs into
protoplanetary disks; for example, X-ray ionization appears likely
to induce magnetodynamical turbulence and MeV flare particles may
produce isotopic anomalies in disk solids \citep{Glassgold00,
Feigelson05a}.

\item The X-ray spectrum gives a measure of the intervening total 
(i.e.  solid and gaseous molecular, atomic, and partially ionized) 
interstellar column density that is derived independently of the 
usual optical-infrared dust extinction curve. The comparison of 
$N_H$ derived from soft X-ray absorption and $A_V$ derived from 
optical-infrared photometry thus allows study of the gas-to-dust 
ratio in molecular clouds and related issues such as interstellar 
metallicities and depletions onto grain surfaces \citep{Vuong03}. 

\item X-ray studies of high-mass O stars elucidate shocks that
thermalize the powerful kinetic energy of their winds.  These
include small-scale shocks in the radiatively accelerated zone near 
the stellar surface (Lucy \& White 1980), intermediate-scale shocks 
arising from magnetic collimation of the wind \citep{Babel97}, and 
parsec-scale shocks when winds collide in rich cluster environments 
\citep{Townsley03}.  Colliding winds from close O star or Wolf-Rayet 
binaries also produce intense penetrating X-ray emission 
\citep{Stevens92}. 

\item By tracing both shocked OB wind flows and supernova remnant
ejection in massive star formation complexes, X-ray images have the
potential to help understand these important paths of energy input
and chemical enrichment of the interstellar medium, providing a
foundation for studies of starburst galaxies
\citep[e.g.][]{Townsley03, Muno04}.

\end{enumerate}

To facilitate advances in these areas, the present paper has two
related goals.  First, we establish the global X-ray properties of
the nearest rich YSC:  the Orion Nebula Cluster (ONC), and its star
formation environment.  The ONC is the only rich cluster with
sufficient complementary optical and near-infrared (ONIR) data to
clearly define the X-ray properties across the IMF.  Second, we
characterize the Orion Nebula as a template for understanding less
sensitive X-ray observations of more distant YSCs.  The $Chandra$
and $XMM-Newton$ satellites have observed over two dozen YSCs and
star formation regions at distances ranging from 0.1 to 10 kpc
during their first five years.

\section{Preparatory material \label{prep.sec}}

Our study is based on the deepest X-ray observation of the Orion 
Nebula nicknamed the $Chandra$ Orion Ultradeep Project (COUP), a 
nearly-continuous 9.7 day exposure of the nebula obtained in January 
2003.  \citet{Getman05a} give a detailed description of the 
observation, data reduction, spectral analysis, source lists and 
characteristics. Most of the data analysis is performed using the 
{\it ACIS Extract} software package\footnote 
{\url{http://www.astro.psu.edu/xray/docs/TARA/ae\_users\_guide.html}}.  
We also rely heavily on the COUP membership study by 
\citet{Getman05b} and analysis of the unobscured optical sample of 
the ONC described by \citet{Preibisch05a}. The reader is referred to 
these studies for background on the sample selection and data 
analysis procedures. A distance of 450 pc to the Orion Nebula is 
assumed throughout these studies.

\subsection{Stellar samples \label{samples.sec}}

We define here a hierarchy of X-ray emitting samples in the Orion
Nebula based on the analyses and tables of \citet{Getman05a,
Getman05b}. These samples are not a complete census of the Orion
Nebula stellar population because many ONIR members are undetected
by COUP \citep[e.g., Tables 11-12 of][]{Getman05a} and are omitted.
The missing stars are mainly very low mass ($M \la 0.2$ M$_\odot$)
stars and brown dwarfs, and some deeply obscured stars.

\begin{description}

\item [Orion Nebula member sample] These are the 1408 COUP sources
which, to the best of our knowledge, constitute all COUP sources
associated with the Orion Nebula star forming region.  This is the
full 1616 COUP sample omitting 208 sources found by \citet{Getman05b}
to be probable non-members:  159 sources without stellar counterparts
classified as `EG' (extragalactic sources), 16 sources with
unobscured stellar counterparts classified as `Field' (foreground
stars), and 33 very faint sources without counterparts classified as
`Unc' (source existence is uncertain).  Forty-two sources classified
as `OMC' and 33 sources classified as `OMC or EG?'  are considered to
be obscured members of the Orion Molecular Cloud.  This can be
considered a complete flux limited sample of Orion Nebula stars in
the $Chandra$ $0.5-8$ keV band.

\item [Hot and cool star subsamples] We divide the Orion Nebula
members into a small group of 10 unobscured hot stars with spectral
types earlier than B4 (B3 to O7), and a large group of 1380 sources
cooler than B4.  The 10 hot `Trapezium' sources (COUP 349, 745, 766,
778, 809, 869, 1116, 1232, 1360, and 1468) are those where the X-ray
emission is likely dominated by stellar wind shock processes,
although contributions by magnetically active cool companions are
sometimes present \citep{Stelzer05}.  Note that there are a handful
of heavily obscured hot stars in the OMC-1, such as the
Becklin-Neugebauer Object (COUP 599b) and Source i (COUP 590), which
may have spectral type earlier than B4 \citep{Grosso05}.  Because
these stars have not been well-characterized yet, we group them with
the cool star subsample.

\item [Lightly and heavily obscured subsamples] Our lightly
obscured subsample consists of 839 cool (spectral type later than
B4) COUP sources of which 822 have absorbing column densities $\log
N_H < 22.0$ cm$^{-2}$ derived from {\it XSPEC} spectral fits
\citep{Getman05a}.  Seventeen sources which are too faint for
spectral fitting with median energies $MedE < 2.0$ keV are added
based on the strong correlation between $\log N_H$ and $MedE$ (\S
\ref{absorb.sec}).  This sample is a superset of the
well-characterized `optical sample' and `lightly absorbed optical
sample' studied by \citet{Preibisch05a} and \citet{Preibisch05b},
respectively. The obscured sample has 559 sources, 541 with $\log
N_H \geq 22.0$ cm$^{-2}$ derived from {\it XSPEC} fits and 18 faint
sources with $MedE \geq 2.0$ keV.  Two groups of these sources
associated with the BN/KL and OMC 1-South cloud cores are discussed 
in detail by \citet{Grosso05}.

\end{description}

The COUP X-ray survey has strengths and weaknesses complementary to
ONIR samples.  It does not suffer confusion from the bright
inhomogeneous nebular emission, it is virtually complete over the 
IMF down to $M \simeq 0.2$ M$_\odot$, and it has a greater ability 
to detect lower mass stars at high obscuration. However, much of the 
$M < 0.1$ M$_\odot$ population is missed \citep{Preibisch05a, 
Preibisch05c} and the sensitivity within $\sim 2$\arcmin\/ of 
$\theta^1$ Ori C is reduced due to contamination by the point spread 
function wings of this extremely X-ray-bright star.

\subsection{X-ray luminosities \label{X_lum.sec}}

We recapitulate the discussion in \S~9 of \citet{Getman05a} on the
reliability and utility of different COUP X-ray luminosity
estimates.  Five X-ray luminosities are considered here: observed
$\log L_s$ in the soft ($0.5-2$ keV) band, observed $\log L_h$ in the
hard ($2-8$ keV) band, intrinsic $\log L_{h,c}$ corrected for X-ray
absorption, observed $\log L_t$ in the total ($0.5-8$ keV) band, and
intrinsic $\log L_{t,c}$.  The absorption corrections are based on
gas column densities $\log N_H$ derived from the X-ray spectra.
Procedures of the spectral modeling and luminosity determinations
are given in \S\S 6-8 of Getman et al.

The five X-ray luminosities are effective for different purposes.
The soft band emission encompasses over half of the intrinsic
emission of cool T Tauri stars and three-quarters of the emission of
hot OB stars.  It is readily compared to measurements made with the
older $Einstein$ $Observatory$ and $ROSAT$ telescopes. But the soft
emission is easily absorbed by intervening interstellar gas; for
typical PMS X-ray spectra, $>90$\% of $\log L_s$ is extinguished 
when $\log N_H > 22.5$ cm$^{-1}$ ($A_V > 30$). Although we could 
operationally correct for this extinction and derive $\log L_{s,c}$ 
values for each star, we do not list them because they can be very 
uncertain. For distant YSCs which must suffer intervening 
absorption, we believe soft band luminosities will not provide 
useful measures of intrinsic X-ray emission.

The hard band is much less vulnerable to extinction.  The
absorption-corrected hard-band luminosities $\log L_{h,c}$ are, we
suggest, the most reliable measure of the intrinsic stellar emission
that can obtain from $Chandra$ data for distant or absorbed young
stellar populations.

The total band luminosities $\log L_t$ and $\log L_{t,c}$ represent
the $Chandra$ data to the fullest extent and provide the strongest
signal.  $\log L_{t,c}$ is the most valuable quantity for
comparisons between lightly obscured YSCs, but it will underestimate
the true emission for obscured or embedded populations due to the
problematic absorption in the soft band. We also recognize that some
of the $\log N_H$ values derived in \citet{Getman05a} may be
inaccurate due to incorrect choice of spectral model (cf. the
outliers in Figure \ref{NH_AV.fig} below), leading to systematic 
errors in $\log L_{t,c}$ values. However, $\log L_{h,c}$ values are 
relatively unaffected by this problem.

\section{Morphology of the Orion Nebula stellar population}

\begin{figure}
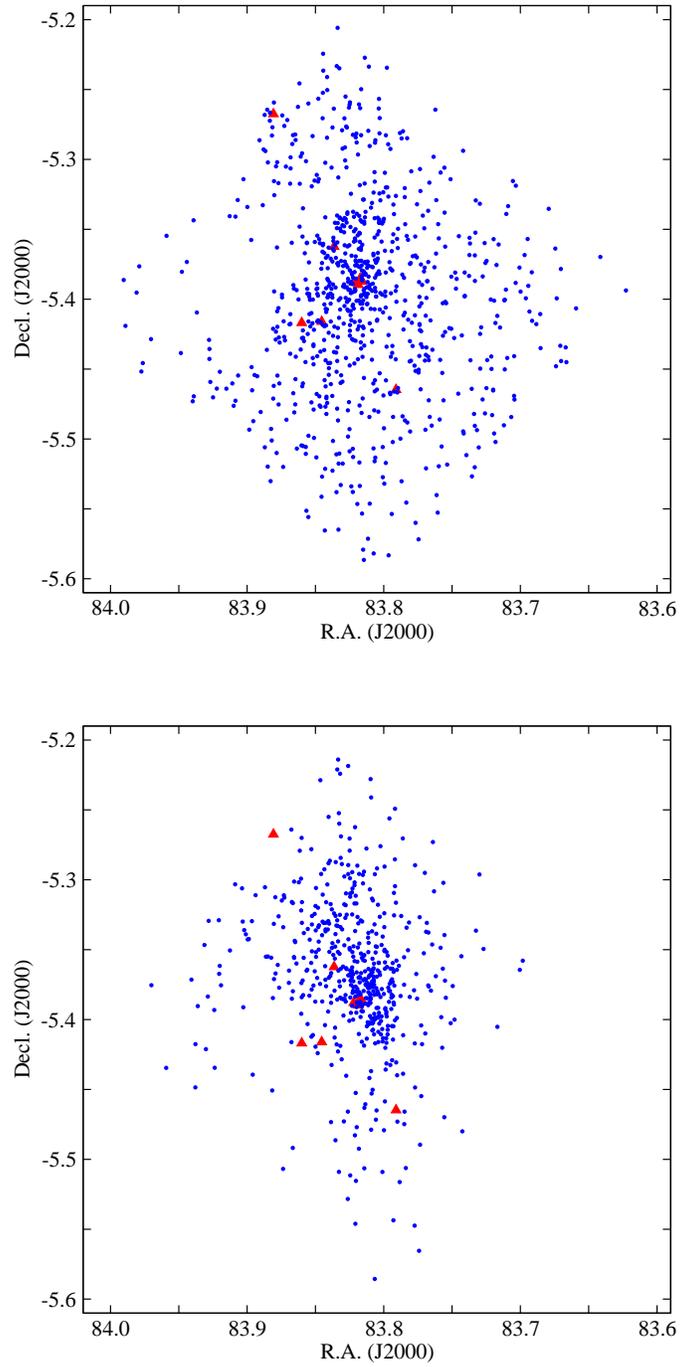

\centering
\begin{minipage}[t]{1.0\textwidth}
  \centering
  \includegraphics[height=0.4\textheight]{f1a.eps}
\end{minipage} \\ [0.4in]
\begin{minipage}[t]{1.0\textwidth}
  \centering
\includegraphics[height=0.4\textheight]{f1b.eps}
\caption{Diagram of the Orion Nebula field showing 1408 COUP X-ray 
sources associated with the Orion Nebula. Top: Lightly obscured 
subsample with $\log N_H < 22.0$ cm$^{-2}$.  Bottom: Heavily 
absorbed subsample with $\log N_H > 22.0$ cm$^{-2}$.  The large 
triangles show the 10 hot O7-B3 stars, while the dots show the 
remaining cool member population.  \label{morph_img.fig}}
\end{minipage}
\end{figure}

The COUP census of the young population provides a new opportunity
to investigate the spatial structure of the Orion Nebula Cluster
which allows inferences regarding the dynamical state of the
cluster.  Figure \ref{morph_img.fig} shows the spatial distribution
of the samples described in \S \ref{samples.sec}.

\subsection{The lightly obscured Orion Nebula Cluster population}

\begin{figure}
\centering
\includegraphics[width=0.9\textwidth]{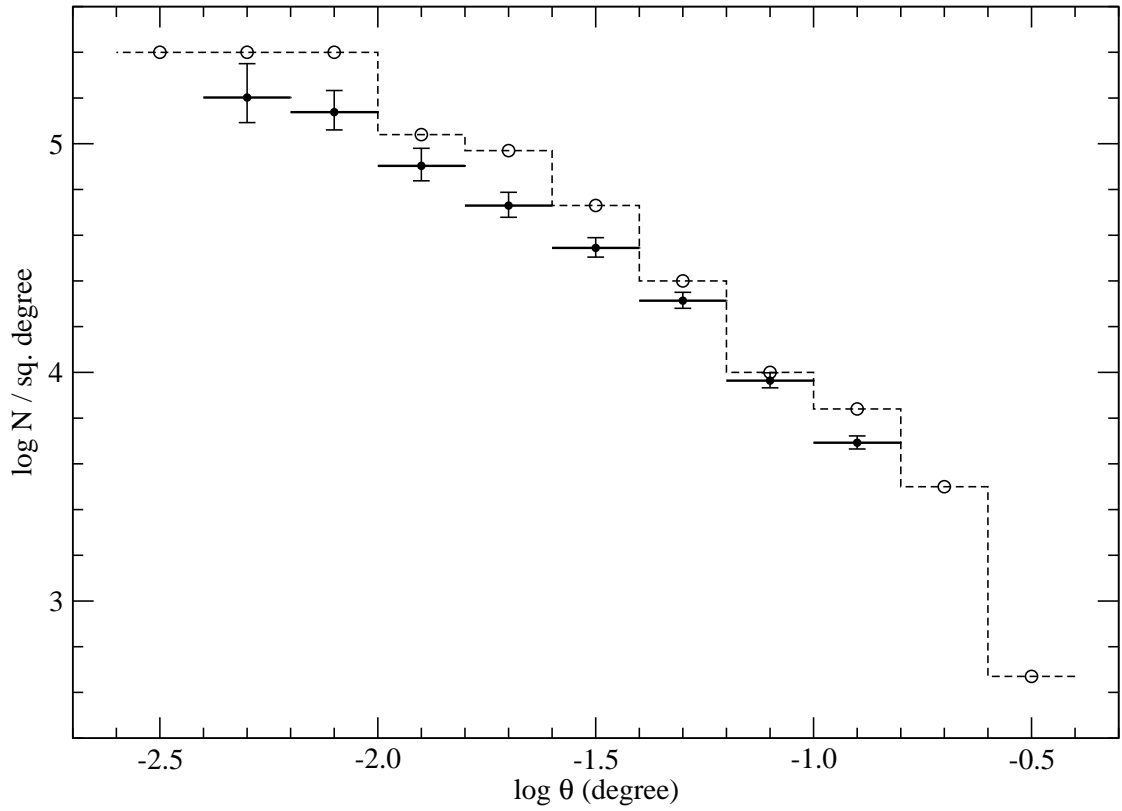}
\caption{Histogram showing the radial profiles of lightly-absorbed
ONC COUP source surface density (disjoint histogram with error bars) 
in comparison with ONIR star surface density (dashed histogram with 
open circles) from \citet{Hillenbrand98}.  \label{morph_rad.fig}}
\end{figure}

The structure of the COUP lightly obscured cool star sample (Figure 
\ref{morph_img.fig} top) follows closely the shape of the ONIR 
sample studied by \citet{Hillenbrand98}, though with about half of 
the stellar density.  This is shown in the comparison of the stellar 
surface density radial profiles plotted in Figure 
\ref{morph_rad.fig}. Their ONIR sample is richer than the COUP ONC 
sample due to their better sensitivity to low-mass members with $M 
\la 0.1$ M$_\odot$, their inclusion of some absorbed stars which we 
place in the heavily absorbed sample, and the reduced sensitivity of 
COUP within 1\arcmin\/ ($\log \theta \leq -1.8$ deg) of $\theta^1$ 
Ori C due to the extended wings of its point spread function.

\subsubsection{The core region}

If one assumes spherical symmetry (but see \S \ref{ONC_asym.sec}),
these radial distributions can be roughly modelled as isothermal
spheres with core radii $r_c \simeq 1$\arcmin\/ or $\sim 0.15$ pc
\citep{Hillenbrand98, Bate98}.  To assist interpretation of other
young stellar clusters, we calculate the X-ray properties of this
$\sim 0.3$ pc diameter core region based on the lightly obscured 
sample.  The central surface density of the cool COUP stars is $\log 
N \sim 5.3$ stars deg$^{-2}$ or $N \sim 2.3 \times 10^3$ stars 
pc$^{-2}$.  Their integrated X-ray luminosities are $\log L_s=32.6$ 
erg s$^{-1}$, $\log L_h$ ($\log L_{h,c}$)~=~32.4 (32.5) erg 
s$^{-1}$, and $\log L_t$ ($\log L_{t,c}$)~=~32.8 (33.0) erg 
s$^{-1}$.  The X-ray surface brightness averaged over the 
core for these cool stars is $5.4 \times 10^{33}$ erg s$^{-1}$ 
pc$^{-2}$, 3.6~(4.4)$\times 10^{33}$ erg s$^{-1}$ pc$^{-2}$, and 
0.9~(1.5)$\times 10^{34}$ erg s$^{-1}$ pc$^{-2}$ in the soft, hard 
and total (absorption-corrected) bands, respectively.  If one 
includes the hot stars lying in the core region ($\theta^1$ Ori C, 
$\theta^1$ Ori A, $\theta^1$ Ori D, and the two COUP components of 
$\theta^1$ Ori B), these central surface brightnesses rise 
to $2.3 \times 10^{34}$ erg s$^{-1}$ pc$^{-2}$, $1 \times 10^{34}$ 
erg s$^{-1}$ pc$^{-2}$, and 3~(5)$\times 10^{34}$ erg s$^{-1}$ 
pc$^{-2}$.  Though few in number, the hot O7-B3 stars dominate the 
X-ray emission in all bands in this central region.

Illustrating the application of these results, we consider the cool
star component of the ONC viewed from a distance of 2 kpc (8 kpc).
Its core region would then subtend 31\arcsec\/ (8\arcsec) with a hard
band flux $F_h \simeq 1 \times 10^{-12}$ erg s$^{-1}$ cm$^{-2}$ ($8
\times 10^{-14}$ erg s$^{-1}$ cm$^{-2}$) producing $\simeq 8300$
($\simeq 500$) counts in a 100 ks $Chandra$ ACIS exposure.  These
hard band count rates are unchanged for most realistic lines-of-sight
through the Galaxy ($\log N_H \la 23$ cm$^{-2}$).  If the point
source detection limit was 10 counts and the cluster was at 2 kpc, 47
stars with $\log L_h \geq 29.9$ erg s$^{-1}$ would be individually
resolved and the integrated emission of 55 fainter stars would
produce a diffuse glow.  If the cluster was at 8 kpc, 8 stars with
$\log L_h \geq 31.1$ erg s$^{-1}$ would be individually resolved.
The integrated hot star contribution would be 2.3-2.5 times higher
than the cool star component and many, but not all, of the stars
earlier than B4 would be individually identified as bright point
sources.  We conclude that the core region of the ONC can be readily
detected and discriminated from point-like sources across the
Galactic disk in {\it Chandra} images.

\subsubsection{Asymmetry in the outer region \label{ONC_asym.sec}}

\citet{Hillenbrand98} model the stellar surface density as ellipses
with ellipticity $\epsilon \simeq 0.3$ oriented nearly north-south.
However, examination of the COUP distribution in Figure
\ref{morph_img.fig} shows a distinct asymmetry: the surface density
of cluster members $3\arcmin - 11\arcmin$\/ east of $\theta^1$ Ori C
is considerably below that seen in a similar region to the west.
This deficit of ONC stars west of $\theta^1$ Ori C is also evident
in optical samples (e.g., Figure~3 of Hillenbrand 1997 and Figure~16
of Bate et al.\ 1998).  From COUP data, we can state with some 
confidence that this deficit is not due to absorption by cloud 
material along the line-of-sight.  An absorbing cloud component 
known as the Dark Bay obscures the H{\sc II} region on the western 
side starting $\sim 1$\arcmin\/ west of $\theta^1$ Ori C 
\citep{ODell01}.  But optical extinction and radio H{\sc I} 
measurements indicate the Dark Bay column density to the bright 
nebula is only $N_H \simeq 3 \times 10^{21}$ cm$^{-2}$ 
\citep{vanderWerf89}, which has negligible effect on the $Chandra$ 
count rates and is insufficient to produce a significant loss in 
sensitivity to X-ray stars.

The east-west asymmetry in the unobscured COUP population thus
appears to be an intrinsic characteristic of the cluster.  The
application of equilibrium dynamical models, such as isothermal
King models, may therefore not be justified, at least for the
outer regions of the cluster on scales $\geq 4$\arcmin\/ or 0.5
pc.  This agrees with the calculation of \citet{Hillenbrand98}
that the cluster has not had time to achieve dynamical
equilibrium; they estimate the two-body encounter relaxation time
is 6.5 Myr for the inner 0.8 pc radius region.  They then suggest
that the cluster is either in the process of violent relaxation
when large-scale asymmetries are prevalent, or its structure today
represents its original configuration at formation.  Based on the
measured velocity dispersion $<\sigma> \simeq 2.5$ km s$^{-1}$,
the crossing time for a typical star across a 0.8 pc distance is
only 0.3 Myr, considerably shorter than the typical $1-2$ Myr age
of ONC stars.  Primordial asymmetries and substructures should
therefore have been quickly erased, as established by 
\citet{Bate98, Scally02}.

We thus conclude that the most likely explanation for the asymmetry
seen in the COUP unobscured population is that the ONC is in the
process of violent relaxation leading to a temporary asymmetrical
spatial distribution of stars \citep{LyndenBell67}.  To our 
knowledge, the ONC is the only case where violent relaxation has 
been arguably detected in a young stellar cluster.

\subsection{The heavily obscured Orion Nebula population}

The large-scale distribution ($>0.5$ pc diameter) of obscured stars 
is more centrally condensed than the lightly obscured ONC component 
and also clearly does not show a dynamically relaxed structure 
(Figure \ref{morph_img.fig} bottom).  These COUP sources tend to lie 
along the north-south dense molecular filament extending from 
several arcminutes south of OMC-1S northward towards OMC-2/3/4 
\citep{Johnstone99}.  A strong concentration of sources is seen 
around the BN/KL region $\sim 1$\arcmin\/ to the northwest of the 
Trapezium, and a second weaker concentration can be seen around the 
Orion 1-South cloud core $1\arcmin-2$\arcmin\/ southwest of 
$\theta^1$ Ori C.  Very similar concentrations of were found among 
$K-L>1.5$ ($A_V \ga 20$) in the sensitive $L$-band study of 
\citet{Lada04}.  The BN/KL and OMC-1S COUP populations are discussed 
in detail by \citet{Grosso05}; the reader should consult this study 
for detailed study of their X-ray properties. 

The existence of such stellar concentrations on $0.1$ pc scales, and 
their associations with molecular cloud cores, is a clear indication 
that these stars are very young.  A population born with $< 
\sigma > = 2.5$ km s$^{-1}$ velocity dispersion would spread $\simeq 
0.5$ pc (4\arcmin) in 0.2 Myr.  The small-scale inhomogeneities 
associated with BN/KL and Orion 1-South thus imply stellar ages of 
$t < 0.1~(\sigma/{\rm{km~s^{-1}}})$ Myr.

It is interesting to note that these structures have not been
clearly evident in earlier studies.  Past ONIR samples have not
penetrated sufficiently deeply into the cloud to match the COUP
obscured sample, where the average absorption is $< \log N_H > =
22.5$ cm$^{-2}$ equivalent to $A_V \simeq 20$ mag.  Past
mid-infrared surveys have been restricted to small regions around
the dense molecular cores, although forthcoming wide-field surveys
with the {\it Spitzer Space Telescope} will overcome this
limitation.

\section{X-ray luminosity function of the Orion population}

\subsection{Global XLF}

Table \ref{Xlum.tab} provides the per-star average and integrated
sample luminosities of the COUP samples in the X-ray bands discussed
in \S \ref{X_lum.sec}.  In the hard band, the integrated
contributions of the lightly and heavily obscured cool star samples
to the integrated Orion Nebula population luminosity are similar
with $\Sigma \log L_{h,c} \simeq 32.8$ erg s$^{-1}$.  The hot stars,
dominated by the uniquely luminous star $\theta^1$ Ori C, emit at
nearly the same level with $\log L_{h,c} = 32.6$ erg s$^{-1}$ (Table
\ref{Xmass.tab}).  Thus, in the hard band which is largely
unaffected by obscuration, the Orion Nebula emission is roughly
equally divided between the lightly obscured ONC, the heavily
obscured stars (including both obscured ONC stars and the embedded
population around the OMC-1 cores), and the few O stars.  Other star
forming regions will necessarily exhibit different ratios between
the unobscured, obscured and OB populations in the X-ray band.  We
suspect that the O stars will dominate the total hard-band emission
in richer YSCs.

\begin{figure}
\centering
\begin{minipage}[t]{1.0\textwidth}
  \centering
  \includegraphics[height=0.4\textheight]{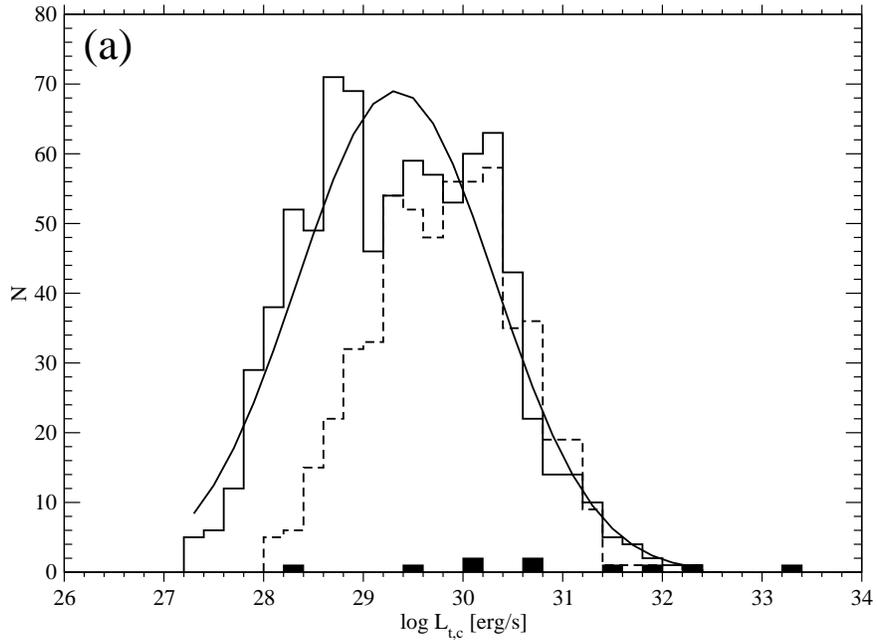} \vspace{0.4in}
\end{minipage} \\ [0.2in]
\begin{minipage}[t]{1.0\textwidth}
  \centering
\includegraphics[height=0.4\textheight]{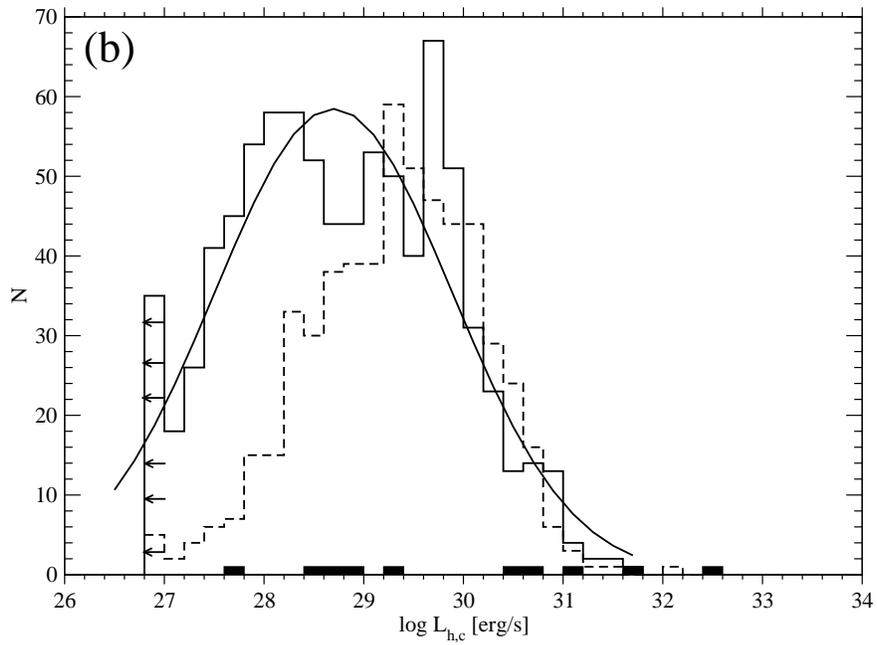}
\caption{Histograms showing the differential distributions (a) $\log 
L_{t,c}$ and (b) $\log L_{h,c}$ for the COUP Orion Nebula 
population.  Solid lines denote the unobscured cool star sample with 
Gaussian fits, dashed lines show the obscured sample, and the 
black-filled histogram show the hot stars including $\theta^1$ Ori 
C. \label{XLF_samp.fig}}
\end{minipage}
\end{figure}

Figure \ref{XLF_samp.fig} shows the binned differential XLF for the 
samples in the total and hard spectral bands, corrected for 
absorption.  Recall that the total band luminosities will 
underestimate true values for heavily absorbed stars, even with the 
attempted correction, because soft components are sometimes 
completely missing from the spectral model.  The shape of the XLF 
can be roughly understood as a convolution of the stellar IMF, which 
breaks from the Salpeter powerlaw below $\simeq 0.5$\,M$_\odot$, and 
the correlation between X-ray luminosity and mass.  These two 
effects result in a steep fall-off in the number of fainter X-ray 
stars in a young stellar cluster.  This explains why the factor of 
ten increase in the limiting sensitivity of COUP over the previous 
$Chandra$ observations of the region \citep{Feigelson02} led to only 
a modest increase in the number of detected lightly-absorbed ONC 
stars.

The XLF of Orion Nebula stars spans 5 orders of magnitude, from
low-mass stars at or below the COUP detection limit of $\log L_t
\simeq 27$ erg s$^{-1}$ to $\geq 1$ M$_\odot$ stars attaining $\log 
L_t \simeq 32$ erg s$^{-1}$ at the peaks of flares.  As noted by 
\citet{Feigelson02} and \citet{Feigelson05b}, the shapes of the XLFs 
are roughly lognormal.  Using the unobscured sample as a template 
for cool star YSC populations in general, we find a $<\log L_{t,c}> 
\simeq 29.3$ erg s$^{-1}$ with standard deviation $\pm 1.0$ and, for 
the hard band, $<\log L_{h,c}> \simeq 28.7$ erg s$^{-1}$ with 
standard deviation $\pm 1.2$.  However, the XLF shapes are not 
precisely lognormal; the tails follow a lognormal shape but the peak 
is nearly flat from $\log L_{t,c} \simeq 28.0$ to 30.5 erg s$^{-1}$.  
Magnetic reconnection flares and systematic decline of X-ray 
luminosity with age will flatten the peak of the XLF, but these 
effects are considerably smaller than the factor of $\simeq 300$ seen 
in the COUP XLF \citep{Preibisch05a, Preibisch05b}.

An important corollary of the lognormal shape at the lower tail is
that extraordinarily few stars appear around the COUP sensitivity
limit.  We thus have high signals on nearly all COUP sources. This 
will not be true for most other $Chandra$ YSC observations where the 
exposure is shorter and the distance is greater.  If the lognormal 
shape at the upper tail is universal, then $Chandra$ observations 
should attempt to reach $\log L_{h,c} \simeq 30.0$ erg s$^{-1}$ to 
capture sufficient cool stars to adequately define the amplitude of 
the curve.  This would provide a powerful means for estimating the 
total YSC population independent of the distribution of OB stars.  
In addition,fitting XLFs to the Orion template offers a new distance 
measure to YSCs across the Galactic disk \citep{Feigelson05b}.

The shapes of the high-luminosity portions (above $\simeq 10^{29}$ 
erg s$^{-1}$ ) of the unobscured and obscured XLFs are very similar, 
but the obscured sample appears to be missing a substantial fraction 
of the weaker sources.  This is true even in the hard band where 
X-rays can penetrate $\log N_H \simeq 23-24$ cm$^{-2}$.  One 
possibility is that a significant fraction of the lower-mass, 
X-ray-weaker sources are obscured by very high column densities, 
perhaps due to localized circumstellar material (i.e. a protostellar 
envelope or disk).  For example, a source with $\log L_{h,c} = 29.0$ 
erg s$^{-1}$ and a typical $kT = 2$ keV thermal spectrum will be 
reduced below the COUP detection limit if $\log N_H \ga 23.7$ 
cm$^{-1}$ ($A_V \ga 300$ mag).

\subsection{Mass-stratified XLF \label{mass.sec}}

Information on masses of Orion Nebula population stars is mostly
restricted to the lightly obscured ONC population with spectroscopic
study by \citet[][updated in Getman et al.  2005a]{Hillenbrand97}.
We examine here 523 COUP sources associated with stars with
spectroscopic mass estimates; this sample is somewhat larger than
the 481 stars considered by \citet{Preibisch05b} because they
truncate their sample at $M < 2$ M$_\odot$.  We reiterate that this
COUP sample is not a complete representation of the underlying
cluster because a large fraction of very low mass cluster members
are missing.  These constitute an important fraction of the cluster
IMF, but collectively contribute very little to the integrated X-ray
luminosity.

\begin{figure}
\centering
\includegraphics[height=0.4\textheight]{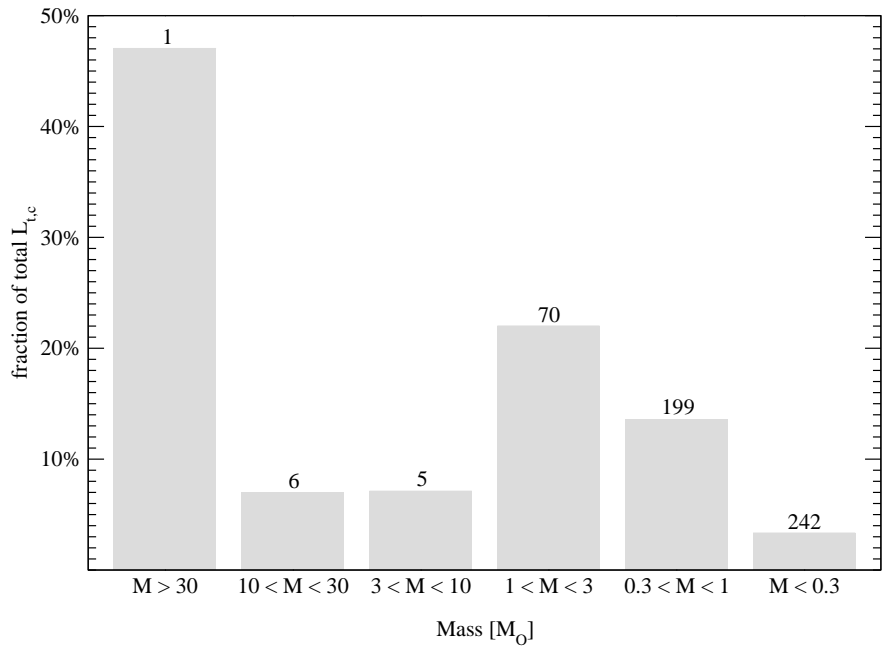}
\caption{Percent of total band ($0.5-8$ keV, corrected for
absorption) X-ray luminosity contributed by different mass strata
using the lightly obscured spectroscopically characterized sample of 
Orion Nebula Cluster stars. The number of stars in each stratum is 
indicated. \label{XLF_fraction mass.fig}}
\end{figure}

\begin{figure}
\centering
\begin{minipage}[t]{1.0\textwidth}
  \centering
  \includegraphics[height=0.4\textheight]{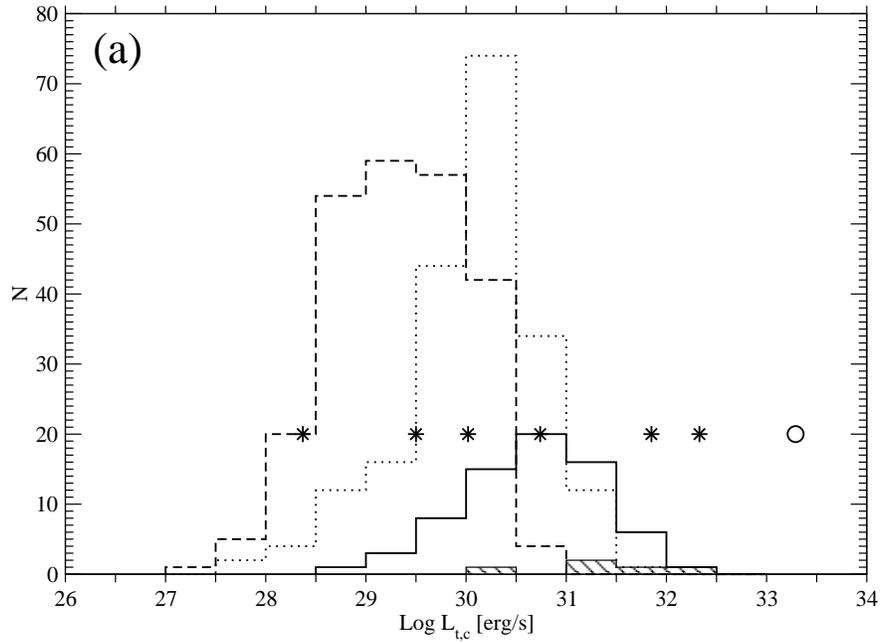} \vspace{0.4in}
\end{minipage} \\ [0.2in]
\begin{minipage}[t]{1.0\textwidth}
  \centering
\includegraphics[height=0.4\textheight]{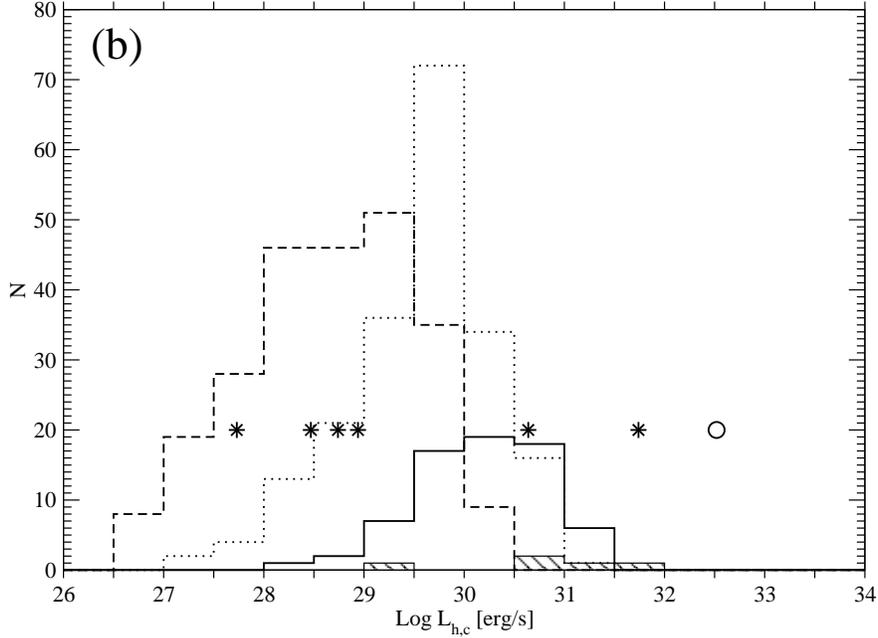}
\caption{Histograms of the distributions (a) $\log L_{t,c}$ and (b) 
$\log L_{h,c}$ for the Orion Nebula Cluster population separated by 
mass strata:  dashed line = $M<0.3$, dotted line = $0.3<M<1$, solid 
line = $1<M<3$, hatched shading = $3<M<10$. Asterisks mark high-mass 
stars with 10 $<$ M $<$ 30 and and the circle marks $\theta^1$ Ori 
C. \label{XLF_mass.fig}}
\end{minipage}
\end{figure}

Table \ref{Xmass.tab} gives the average and integrated luminosities 
of mass ranges in the ONC.  Figure \ref{XLF_fraction mass.fig} 
summarizes the results by showing the percent of the total emission 
$L_{t,c}$ contributed by different mass strata. Figure 
\ref{XLF_mass.fig} shows the mass-stratified differential XLFs.  The 
table shows that 54\% (34\%) [47\%] of the total nebula luminosity 
in the soft (hard) [absorption-corrected total] band is produced by 
the most massive $M>30$ M$_\odot$ member, $\theta^1$ Ori C.  
Omitting $\theta^1$ Ori C from further consideration, Figure 
\ref{XLF_mass.fig} show that the solar-mass stratum ($1-3$ 
M$_\odot$) dominates the integrated cluster X-ray luminosity, 
contributing about 41\% (41\%) [41\%] in the soft (hard) [total] 
band.  Stars around the peak of the IMF with masses $0.3-1$ 
M$_\odot$ are individually fainter but more numerous.  They 
collectively contribute 19\% (32\%) [26\%] of the cluster 
luminosity.  The very low mass stars and brown dwarfs contribute 
less than 7\% of the ONC emission in all bands.

A clear trend of X-ray luminosities increasing with mass is seen in 
the XLFs (Figure \ref{XLF_mass.fig}).  The $L_x-M$ trend was 
suggested in $ROSAT$ data \citep{Feigelson93} and is well-documented 
in $Chandra$ studies of the ONC \citep{Feigelson03, Flaccomio03, 
Preibisch05a}.  Note that the $10<M<30$ M$_\odot$ subsample does not 
obviously follow the trend. \citet{Stelzer05} find that the ONC B 
star population does not have homogeneous X-ray properties:  some 
are X-ray quiet, others have magnetically confined wind emission, 
and others emit over a wide range of luminosities due to unresolved 
lower mass companions.

Detection of individual sources in $Chandra$ observations of distant 
YSCs will be flux limited at some level in the XLFs shown in Figure 
\ref{XLF_mass.fig} depending on the cluster distance $d$ and the 
duration of the exposure $t_{exp}$.  Note that $Chandra$ 
sensitivities increase linearly with exposure time due to the low 
instrumental background, unless a bright diffuse X-ray component is 
present.  A rough estimate of the on-axis $Chandra$ ACIS point 
source sensitivity limit based on COUP is $$ \log L_{h,c,lim} \simeq 
28.7 + 2 \log (d/\rm{kpc}) - \log (t_{exp}/100~\rm{ks}) + 0.25 (\log 
N_H - 20.0)~ \rm{erg~ s^{-1}}.$$ The absorption dependence is 
obtained using the PIMMS software assuming a thermal plasma with $kT 
= 2$ keV.  For the total band, the dependence on absorption is $0.4 
( \log N_H - 20.0)$ (see equation 3 of Feigelson et al.  2002).

For a short YSC observation with sensitivity limit $\log
L_{h,c,lim} = 31.0$ erg s$^{-1}$, only 15 stars at the
high-luminosity tip of the cluster XLF are sampled (in addition to
some high mass stars that we omit from consideration here) and
these sources will often be detected only briefly near the peak of
flares.  With a sensitivity limit of $\log L_{h,c,lim} = 30.0$ erg
s$^{-1}$, 228 individual sources are detected.  Most of these will
be stars with $M > 0.5$ M$_{\odot}$.  When the sensitivity is
higher with $\log L_{h,c,lim} = 29.0$ erg s$^{-1}$, 729 stars are
detected.  These comprise nearly all cluster members with $M >
1.0$ M$_\odot$ and most of the stars with $0.3 < M < 1.0$
M$_\odot$.  Sensitivities of $\log L_{h,c,lim} = 28.0$ erg
s$^{-1}$ capture 81\% of the COUP population, or 1140 of 1408
stars including nearly all of the cluster population with $M >
0.3$ M$_\odot$.  Only a modest improvement (15\% of member stars)
is made by further increasing the sensitivity down to $\log
L_{h,c,lim} = 27.0$ erg s$^{-1}$ due to the steep drop in the
cluster XLF.

As the sensitivity limit increases from $\log L_{h,c,lim} = 31.0$ to
30.0 (29.0) [28.0] erg s$^{-1}$ as one moves from distant to nearby
YSCs, the integrated luminosity of the cluster stars which are not
individually detected declines from 32.9 to 32.3 (32.2) [30.0] erg
s$^{-1}$.  These photons will appear as a diffuse X-ray structure
around the detected sources.  In lower sensitivity images, the
integrated emission from unresolved YSC stars can be confused with
intrinsically diffuse plasma associated with O star wind shocks.
This issue is discussed by \citet{Townsley03}.

\section{Composite spectrum of ONC stars \label{spec.sec}}

\begin{figure}
\centering
\includegraphics[width=0.9\textwidth]{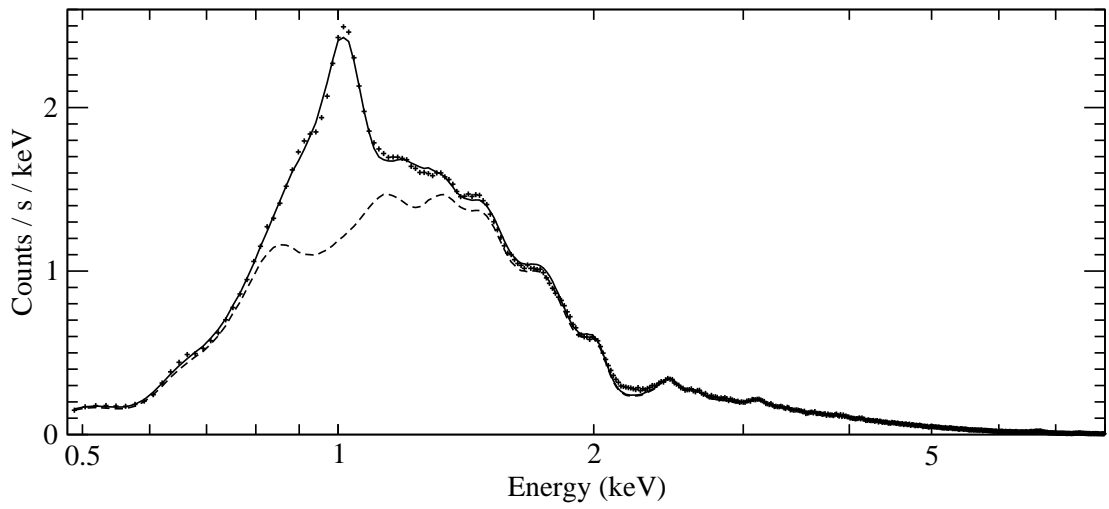}
\caption{Composite spectrum of 820 cool stars without detector 
pileup, constituting 40\% of the Orion cool star lightly absorbed 
X-ray emission. The best fit model (see text for parameters) is 
shown with the solid line. The dashed line is the same model with Ne  
abundance artificially set to solar levels. Note the ordinate is in 
linear units, unlike most COUP spectra shown with logarithmic units. 
\label{comp_spec.fig}}
\end{figure}

The lightly obscured subsample of 839 stars includes 19 which are 
strong enough to suffer photon pileup in the ACIS-I detector. We 
first considered the 820 COUP sources without pileup.  The {\it ACIS 
Extract} software was used to create a composite spectrum and 
exposure-weighted auxiliary response files (ARFs) for spectral 
analysis using the {\it XSPEC} software \citep[for details, see \S 7 
of][]{Getman05a}. The resulting composite spectrum and best-fit 
two-temperature model are shown in Figure \ref{comp_spec.fig}.  The 
model parameters are column density $N_H = 3 \times 10^{21}$ 
cm$^{-2}, $ $kT_{1} = 0.5$ keV, $kT_{2} = 3.3$ keV with non-solar 
abundances.  The integrated luminosity of these 820 stars over the 
total $0.5-8$ keV band is $\log L_{t,c} = 32.82$ erg s$^{-1}$; the 
luminosity of the 19 piled-up stars is $\log L_{t,c} = 33.00$ erg 
s$^{-1}$.  Thus, the spectrum in Figure \ref{comp_spec.fig} 
constitutes about 40\% of the lightly obscured Orion Nebula
stellar X-ray emission.  

Abundance excesses far above (factor of 10 or more) solar were 
found for oxygen, nitrogen, neon, magnesium, sulfur, argon, calcium 
and iron.  The fit is not statistically valid and the reliability of 
the individual abundances is uncertain; this is not surprising given 
the enormous complexity of the plasma from hundreds of magnetically
active stars. However, visual examination of the spectrum in Figure 
\ref{comp_spec.fig} shows that the strongest line complex is from Ne 
IX and Ne X lines around 1 keV\footnote{We also examined the 
composite spectrum of 19 piled-up cool stars, dividing each ARF by its 
point spread function fraction appropriate to the annular extraction 
region. It also shows a strong line complex around 1 keV.}.  The 
dashed spectrum shows the best-fit spectrum with the neon abundance 
artificially set at solar levels.  The flux at 1 keV is effectively 
doubled by the excess neon, which constitutes about 7\% of the total 
$0.5-8$ keV luminosity.  Excess emission at 1 keV thus has the 
potential to be a useful diagnostic of magnetically active stars in 
YSCs, providing they are not so heavily obscured that the 1 keV 
emission is absorbed.

\section{Absorption in the obscured population \label{absorb.sec}}

The absorption of soft X-rays by bound-free electronic transitions
provides a unique way to measure line-of-sight column densities of
interstellar matter.  ONIR extinction curves measure only dust
particles, 21-cm emission measures only warm HI, and molecular line 
studies measure only specific gas-phase species.  In comparison, 
X-ray absorption measures the integrated effects of various elements 
(C, O, N and Ne are most important although H and He play a 
significant role) in all phases of intervening material. The 
spectral dependence of the photoelectric absorption scales as 
(E/keV)$^{-2.65}$ for a solar abundance mixture of elements 
\citep{Wilms00, Glassgold00}, and the integrated effect on the 
intrinsic spectrum is traditionally expressed in terms of the 
equivalent column density $N_H$ of hydrogen atoms along the 
line-of-sight.  The most detailed study to date that compares X- ray 
absorption to ONIR dust extinction used obscured young stars in the 
$\rho$ Ophiuchi cloud to derive the gas-to-dust absorption ratio 
$N_H/A_V = 1.6 \times 10^{21}$ atoms cm$^{-2}$ mag$^{-1}$ 
\citep{Vuong03}. A similar study of obscured COUP sources is 
underway.  Because the absorption varies widely from $\log N_H < 20$ 
to $\simeq 24$ cm$^{-1}$ ($A_V = 0$ to $\simeq 500$ mag) from star 
to star in the COUP observations, and will also differ between YSCs 
depending on their location in the Galaxy and in their natal 
molecular cloud, the distribution of absorptions seen in Orion has 
no direct relevance to studies of other YSCs.

\begin{figure}
\centering
\includegraphics[height=0.6\textheight]{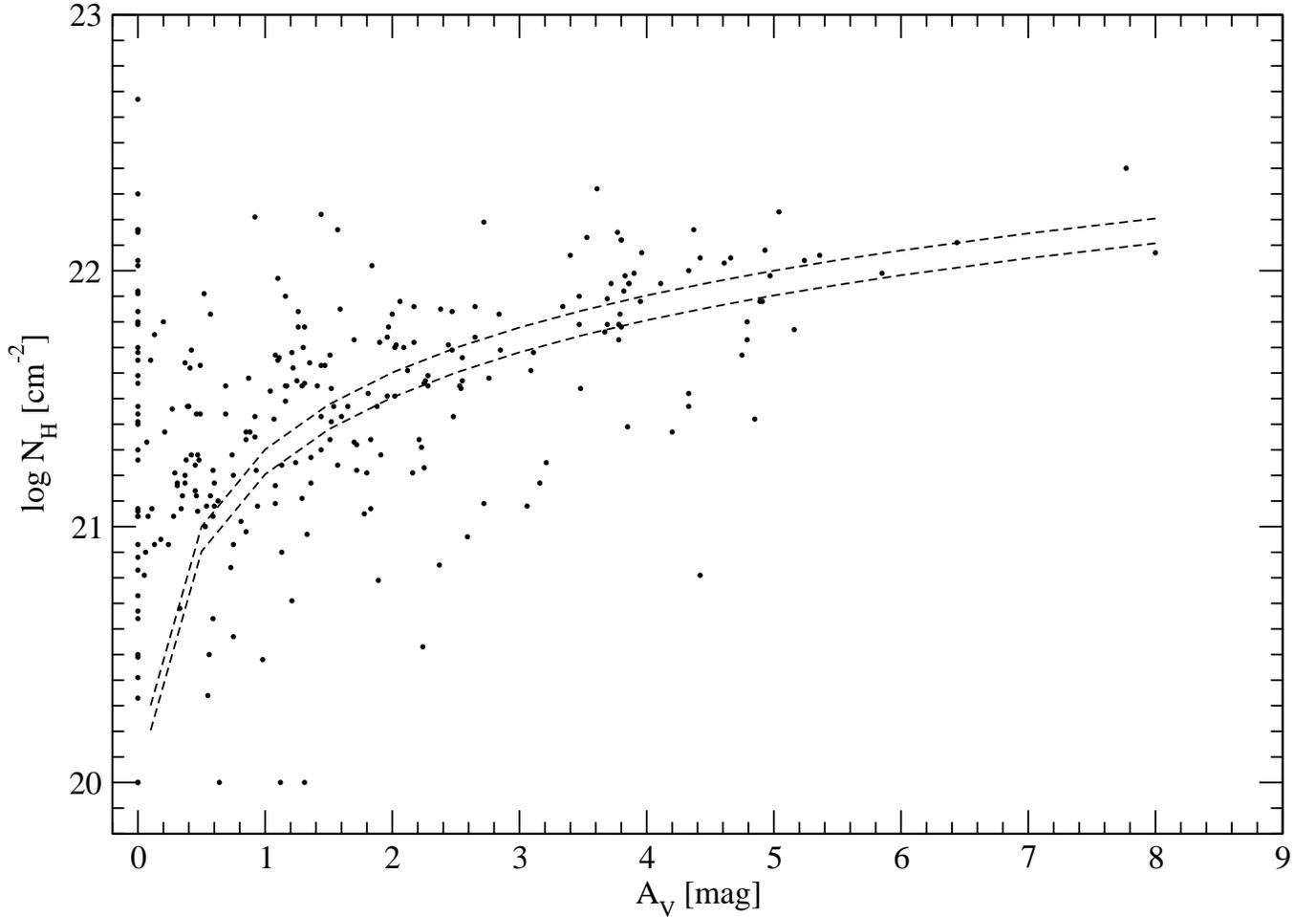}
\caption{Plot of $\log N_H$ (cm$^{-2}$) from COUP spectral fits of 
268 bright ($NetCts> 1000$) X-ray sources and visual absorptions 
$A_V$ (mag) obtained by \citet{Hillenbrand97}. The two curves give 
the gas-to-dust relationship $N_{H} = 1.6 \times 10^{21}A_{V}$ 
(lower) and $N_{H} = 2.2 \times 10^{21}A_{V}$ (upper) derived by 
\citep{Vuong03} and \citep{Ryter96}, respectively. 
\label{NH_AV.fig}}
\end{figure}

Figure \ref{NH_AV.fig} plots $\log N_H$ from COUP {\it XSPEC} fits 
\citep{Getman05a} against dust absorption $A_V$ derived by 
\citet{Hillenbrand97} from optical photometry and spectroscopy.  The 
sample shown is restricted to 268 COUP sources with more than 1000 
source counts\footnote{A similar plot constructed with ten times 
fewer photons for X-ray spectroscopic fitting by \citet{Feigelson02} 
showed many more discrepant points.  Some showed high $A_V$ with 
very low $\log N_H$ while others, as in Figure \ref{NH_AV.fig}, 
showed zero $A_V$ with high $\log N_H$.  Such discrepant points are 
still present in the full COUP sample, but many disappear when the 
COUP data are restricted to sources with $NetCts > 1000$ counts 
where the {\it XSPEC} fitting procedure is most reliable.  The 
formal error on $\log N_H$ from {\it XSPEC} fitting is typically 
$<0.05$.  We believe the scatter and discrepancies in the diagram 
shown here are probably no longer due to {\it XSPEC} fitting 
errors.}.  The COUP $\log N_H$ values generally, but not always, 
agree with the visual absorption for $N_H/A_V \simeq 2 \times 
10^{21}$ atoms cm$^{-2}$ mag$^{-1}$.  The values set at $A_V = 0.0$ 
represent cases where the photometric colors are too blue for the 
spectral type.  

The discrepancies between $log N_H$ and $A_V$ might arise from
instrumental difficulties, such as erroneous choice of X-ray spectral
model (as shown in Figure 6 of Getman et al.\ 2005) or errors in the
optical photometry due to photospheric variability, binarity,
scattered light, or inaccurate calibration of pre-main sequence
photospheric colors.  They may also represent unusual dust properties
in the local environments around certain young stellar objects.  SVS
16 in the NGC~1333 cloud and EC 95 in the Serpens cloud are two clear
case where $A_V$ derived from infrared photometry was overestimated
compared to the X-ray-derived $\log N_H$ \citep{Getman02,
Preibisch03}, perhaps due to unusual geometries of circumstellar dust
or the disappearance of the gaseous component of the disk.  In the
COUP sources with discrepant absorption measurements, $A_V$ may be
too low because circumstellar disk grains have coagulated into large
particles while the gas component is still present.

\begin{figure}
\centering
\includegraphics[height=0.6\textheight]{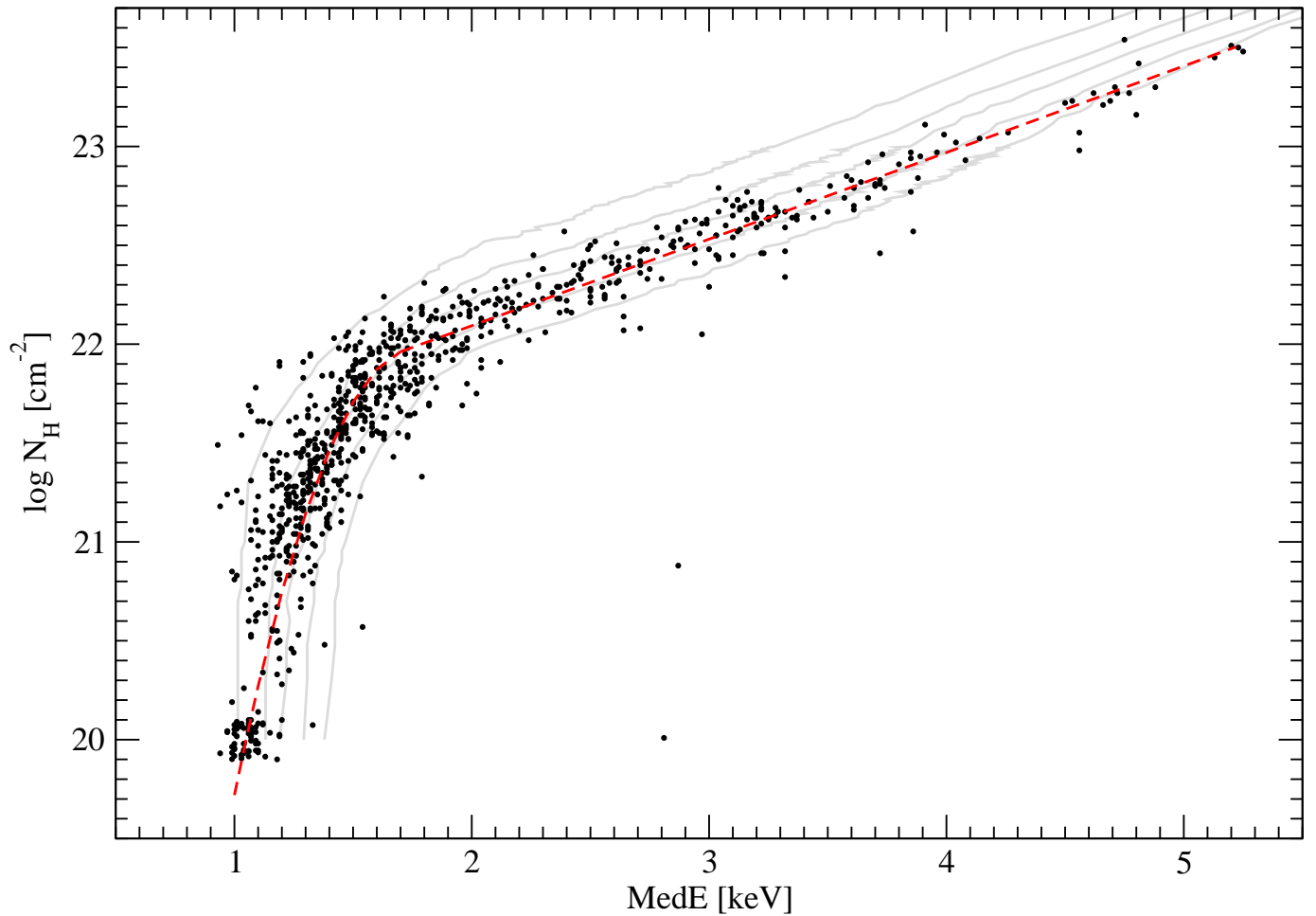}
\caption{ Plot of total column density $\log N_H$ (cm$^{-2}$) and 
median energy of background-subtracted events (black dots). The 
sample consists of 844 COUP members stronger than $NetCts
> 200$ cts.  The red dashed curve is an empirical fit to the data with two
curves: quadratic and linear. Gray lines indicate, from left to 
right, the locus of $kT=(1.0,1.5,2.0,3.0,5.0)$ keV thermal spectra 
predicted from absorbed MEKAL model.\label{NH_MedE.fig}}
\end{figure}

The COUP study can also provide practical guidance for estimation of 
individual $\log N_H$ measurements in YSCs where the signal may be 
insufficient for nonlinear parametric spectral modelling of 
individual sources.  Figure \ref{NH_MedE.fig} shows that the median 
energy of background-subtracted ACIS events for each source, the 
quantity $MedE$ listed in Table 3 of Getman et al.\ (2005), is a 
reliable indicator for the absorbing column density $\log N_H$ 
obtained from spectral fitting procedures. Other researchers have
also found that median energies are effective spectral estimators
in X-ray CCD spectroscopy \citep{Hong04}. The nonlinear 
relationship can be empirically approximated by two 
curves\footnote{The quantitative relationship found here must be 
modified for observations obtained when the soft X-ray absorption of 
contaminating material on the ACIS-I filter differed substantially 
from the 2003 COUP observation. It also must be recalibrated to be 
used for observations taken with different telescopes and detectors, 
such EPIC on {\it XMM-Newton}.}.  For relatively soft sources $MedE 
= [1.0 - 1.7]$ keV, $\log N_H = 9.96 + 13.62 \times MedE - 3.86 
\times MedE^{2}$ cm$^{-2}$.  For harder sources with $MedE > 1.7$ 
keV, $\log N_H = 21.22 + 0.44 \times MedE$ cm$^{-2}$.  A 
concatenation of these curves is plotted as the red dashed line in 
Figure \ref{NH_MedE.fig}. For values of $MedE$ around $1.0-1.3$ keV, 
the scatter is great and one can confidently conclude only that 
$\log N_H < 22$ cm$^{-2}$.  Since $MedE$ is easily obtained for 
sources as faint as a few counts, which often dominate samples in 
distant and obscured YSCs, this empirical conversion from $MedE$ 
provides an effective procedure for approximating $\log N_H$ values 
for individual stars.  Note that any source with $MedE< 1.0$ keV is 
probably too soft to be a member of the low-mass Orion population 
even with no obscuration; such sources may be older foreground field 
disk stars or shocks from protostellar outflows.

Using the gray curves in Figure~\ref{NH_MedE.fig} derived from
{\it XSPEC} simulations of absorbed one-temperature thermal
spectra, we also note a trend of increasing temperature with
increasing median energy.  Plasma temperatures are typical $kT
\sim 1 - 2$ for $MedE < 1.5$, and increase to $kT \sim 2-5$ for
$MedE > 2$.  This trend can further assist in characterizing
distant and faint Orion population stars.

\section{Summary}

This study establishes the global X-ray properties of the stellar
population associated with the Orion Nebula stellar population --
the light-obscured Orion Nebula Cluster (ONC) and the
heavily-obscured stars associated with molecular material in its
vicinity.  The analysis is based on the $Chandra$ Orion Ultradeep
Project's 9.7 day observation of the region in January 2003
described by Getman et al.\ (2005).  Care is taken to remove
extragalactic and foreground contaminants from the sample.  The
information serves both to further understanding of the nearest rich
YSC, and to use the Orion Nebula as a template for understanding
less sensitive X-ray observations of more distant
YSCs\footnote{\citet{Helfand01} perform a related calculation of the
integrated X-ray luminosity arising from OB populations and obtain a
range of luminosity $2-20 \times 10^{34}$ erg s$^{-1}$ per O star in
the $2-10$ keV band.  This is $\sim 20-200$ times larger than the
value we obtain in the Orion Nebula because their result is
dominated by the contributions of supernova remnants and X-ray
binary systems from earlier generations of massive star formation.
The Helfand-Moran value is more appropriate for large-scale and
long-lived starburst complexes and galaxies, while our result is
more applicable to small-scale and short-lived individual YSCs.
Observers of distant high-mass star forming regions must exercise
caution, as supernova remnants and X-ray binaries may be present
near regions of current star formation.}.

The X-ray emission of the Orion Nebula region can be divided into
three components:  several OB stars dominated by $\theta^1$ Ori C, a
rich population of 822 lightly obscured cool stars associated with
the Orion Nebula Cluster, and a population of 559 heavily obscured
stars.  These components contribute roughly equally to the Orion
Nebula stellar emission in the hard $2-8$ keV band; the integrated
absorption-corrected luminosity of each component is about $\log
L_{h,c} \simeq 32.7$ erg s$^{-1}$ in the hard $2-8$ keV band and
$\log L_{t,c} \simeq 33.2$ erg s$^{-1}$ in the total $0.5-8$ keV
band.

The lightly obscured sample shows a radial profile similar to that
seen in optical and near-infrared samples.  The inner region can be 
characterized by an isothermal core emitting $\simeq 1 \times 
10^{34}$ erg s$^{-1}$ pc$^{-2}$ in the $0.5-8$ keV band.  The COUP 
stars in the outer region on scales $0.5-2$~pc from $\theta^1$ Ori C 
show a strong spatial asymmetry that we attribute to a temporary 
period of violent relaxation in the dynamical evolution of the 
cluster. The obscured COUP sources show concentrations around both 
OMC-1 molecular cores.  Their small-scale substructures indicate 
these stars are very young, probably with ages $t \la 0.1$ Myr.

The XLF of the lightly obscured sample is approximately lognormal in 
shape with $<\log L_{t,c}>=29.3$ erg s$^{-1}$ and standard deviation 
$1.0$.  The peak of the XLF shape can serve as a standard candle for 
a new distance measure to YSCs across the Galactic disk.  The 
obscured population shows a deficit in lower-luminosity stars, even 
in the hard band.  This suggests a significant population of stars 
with localized circumstellar material producing column densities 
$\log N_H \ga 24$ cm$^{-2}$ is missing from the COUP sample. 
Mass-stratified XLFs show that, in the hard band, one-third of the 
Orion Nebula emission is produced by the X-ray-bright O6 star 
$\theta^1$ Ori C and half is produced by lower mass pre-main 
sequence stars with masses $0.3<M<3$ M$_\odot$.  The numerous very 
low mass stars and brown dwarfs contribute little to the total X-ray 
emission. 

The integrated spectra of the low mass stars is well fit by a 
two-temperature plasma model with super-solar abundances of several 
elements.  The neon line complex around 1 keV is particularly strong 
and might serve as a tracer of magnetically active stars in distant 
YSCs.   We establish that the median energy of ACIS counts is an 
excellent predictor for the absorbing column density $\log N_H$ in 
Orion stars.  However, discrepancies between $\log N_H$ from gas and 
visual absorption $A_V$ from dust are sometimes present.  The cause 
of this is uncertain.  

Using the $2-8$ keV hard band emission, we show that young stellar 
clusters like the Orion Nebula can be readily detected and resolved 
with $Chandra$ across the Galactic disk, even in the presence of 
heavy obscuration.  This confirms earlier arguments that X-ray 
telescopes can effectively trace star formation throughout the 
Galaxy \citep{Montmerle87, Montmerle02}.  For more distant clusters, 
however, only the brightest X-ray stars are individually detected 
while the remaining population is seen as a diffuse glow. Our study 
of the Orion Nebula as a template for X-ray studies of distant star 
forming regions gives this principle a stronger quantitative 
foundation. 

\acknowledgements

We thank Lynne Hillenbrand (Caltech), Guiseppina Micela (Palermo), 
Alexander Tielens (Groningen) and the referee Robert O'Dell 
(Vanderbilt) for valuable discussions. COUP is supported by 
{\it Chandra} guest observer grant SAO GO3-4009A (E.\  Feigelson, 
PI).  This work was also supported by the ACIS Team contract 
NAS8-38252.


\clearpage
\newpage

\clearpage
\newpage

\clearpage
\newpage

\clearpage
\newpage

\clearpage
\newpage

\clearpage
\newpage

\clearpage
\newpage

\clearpage
\newpage


\begin{deluxetable}{lrrrrr}
\tablecolumns{6} \tablewidth{0pt}

\tablecaption{Global X-ray luminosities of the Orion Nebula stellar 
samples \label{Xlum.tab}}

\tablehead{ \colhead{Quantity} & \colhead{Orion Nebula} & 
\colhead{Lt.\ obscured} & \colhead{Hv.\ obscured} & \colhead{Hot} & 
\colhead{Cool} \\

& \colhead{members} & \colhead{cool stars} & \colhead{stars} & 
\colhead{stars} & \colhead{stars} }

\startdata
COUP cts              &  15,060,010 & 6,160,850 & 1,046,160 & 7,853,000 & 7,207,010 \\
Num stars             &        1408 &       839 &       559 &        10 &      1398 \\
&&&&&\\
\multicolumn{6}{c}{Median luminosity per star} \\
$<\log L_s>$          &        28.51&       28.87&        27.91 &   30.17 & 28.50 \\
$<\log L_h>$          &        28.90&       28.71&        29.09 &   30.41 & 28.90 \\
$<\log L_{h,c}>$      &        29.06&       28.74&        29.36 &   30.42 & 29.06 \\
$<\log L_t>$          &        29.15&       29.13&        29.14 &   30.57 & 29.14\\
$<\log L_{t,c}>$      &        29.59&       29.35&        29.85 &   30.74 & 29.58\\
&&&&&\\
\multicolumn{6}{c}{Integrated luminosity in sample} \\
$\Sigma \log L_s$     &        33.33&       32.90&        31.72 &   33.11 & 32.93\\
$\Sigma \log L_h$     &        33.14&       32.77&        32.61 &   32.60 & 33.00\\
$\Sigma \log L_{h,c}$ &        33.21&       32.79&        32.78 &   32.61 & 33.09\\
$\Sigma \log L_t$     &        33.55&       33.14&        32.66 &   33.22 & 33.27\\
$\Sigma \log L_{t,c}$ &        33.76&       33.28&        33.18 &   33.36 & 33.54\\
\enddata
\end{deluxetable}

\clearpage
\newpage

\begin{deluxetable}{lrrrrrr}
\tablecolumns{7} \tablewidth{0pt}

\tablecaption{Mass-stratified X-ray luminosities of the Orion Nebula
Cluster \label{Xmass.tab}}

\tablehead{ \colhead{Quantity} & \colhead{$M>30$} &
\colhead{$10<M<30$} & \colhead{$3<M<10$} & \colhead{$1<M<3$} &
\colhead{$0.3<M<1$} & \colhead{$M<0.3$} \\

& \colhead{M$_\odot$} & \colhead{M$_\odot$} & \colhead{M$_\odot$} &
\colhead{M$_\odot$} & \colhead{M$_\odot$} & \colhead{M$_\odot$} }

\startdata
COUP cts  &6,344,400  & 1,430,720 & 1,119,870 & 3,149,960 & 1,376,130 & 349,019 \\
Num stars &        1  &         6 &         5 &        70 &       199 &     242 \\
&&&&&&\\
\multicolumn{6}{c}{Median luminosity per star} \\
$<\log L_s>$          & 33.03 & 30.17 & 30.75 & 30.15 & 29.49 & 28.92 \\
$<\log L_h>$          & 32.51 & 28.94 & 30.93 & 30.22 & 29.65 & 28.65 \\
$<\log L_{h,c}>$      & 32.52 & 28.94 & 30.97 & 30.26 & 29.69 & 28.66 \\
$<\log L_t>$          & 33.14 & 30.18 & 31.22 & 30.49 & 29.92 & 29.14 \\
$<\log L_{t,c}>$      & 33.29 & 30.74 & 31.43 & 30.69 & 30.16 & 29.39 \\
&&&&&&\\
\multicolumn{6}{c}{Integrated luminosity in mass stratum} \\
$\Sigma \log L_s$     & 33.03 & 32.31 & 32.17 & 32.60 & 32.26 & 31.66 \\
$\Sigma \log L_h$     & 32.51 & 31.77 & 31.89 & 32.42 & 32.30 & 31.62 \\
$\Sigma \log L_{h,c}$ & 32.52 & 31.77 & 31.90 & 32.44 & 32.33 & 31.65 \\
$\Sigma \log L_t$     & 33.14 & 32.42 & 32.35 & 32.82 & 32.58 & 31.94 \\
$\Sigma \log L_{t,c}$ & 33.29 & 32.46 & 32.47 & 32.96 & 32.75 & 32.14 \\
\enddata
\end{deluxetable}


\clearpage
\newpage

\begin{thebibliography}

\bibitem[Babel \& Montmerle(1997)]{Babel97} Babel, J., \&
Montmerle, T.\ 1997, \apjl, 485, L29
\bibitem[Bate et al.(1998)]{Bate98} Bate, M.~R., Clarke, 
C.~J., \& McCaughrean, M.~J.\ 1998, \mnras, 297, 1163 
\bibitem[Carpenter et al.(2001)]{Carpenter01} Carpenter, J.~M.,
Hillenbrand, L.~A., \& Skrutskie, M.~F.\ 2001, \aj, 121, 3160
\bibitem[Favata \& Micela(2003)]{Favata03} Favata, F., \&
Micela, G.\ 2003, Space Science Reviews, 108, 577
\bibitem[Feigelson et al.(1993)]{Feigelson93} Feigelson, E.~D.,
Casanova, S., Montmerle, T., \& Guibert, J.\ 1993, \apj, 416, 623
\bibitem[Feigelson et al.(2002)]{Feigelson02} Feigelson, E.~D.,
Broos, P., Gaffney, J.~A., Garmire, G., Hillenbrand, L.~A., Pravdo,
S.~H., Townsley, L., \& Tsuboi, Y.\ 2002, \apj, 574, 258
\bibitem[Feigelson \& Montmerle(1999)]{Feigelson99} Feigelson,
E.~D., \& Montmerle, T.\ 1999, \araa, 37, 363
\bibitem[Feigelson et al.(2003)]{Feigelson03} Feigelson, E.~D.,
Lawson, W.~A., \& Garmire, G.~P.\ 2003, \apj, 599, 1207
\bibitem[Feigelson et al.(2004)]{Feigelson04} Feigelson,
E.~D., Hornschemeier, A.\ E., and 7 others  2004, \apj, 611, 1107
\bibitem[Feigelson(2005)]{Feigelson05a} Feigelson, E.\ D.\ 2005,
in Cool Stars, Stellar Systems and the Sun 13 (F.\ Favata \& J.\
Schmitt, eds.), ESA SP, in press
\bibitem[Feigelson \& Getman(2005)]{Feigelson05b} Feigelson, E.\
D., \& Getman, K.\ V.\ 2005, in The Initial Mass Function: Fifty
Years Later (E. Corbelli et al., eds.), Kluwer, in press
\bibitem[Flaccomio et al.(2003)]{Flaccomio03} Flaccomio, E.,
Damiani, F., Micela, G., Sciortino, S., Harnden, F.~R., Murray,
S.~S., \& Wolk, S.~J.\ 2003, \apj, 582, 398
\bibitem[Getman et al.(2002)]{Getman02} Getman, K.~V.,
Feigelson, E.~D., Townsley, L., Bally, J., Lada, C.~J., \& Reipurth,
B.\ 2002, \apj, 575, 354
\bibitem[Getman et al.(2005a)]{Getman05a} Getman, K.,V., Flaccomio,
E. and 22 co-authours\ 2005, \apjs, in press, astro-ph/0410136
\bibitem[Getman et al.(2005b)]{Getman05b} Getman, K.\ V., Feigelson,
E.\ D., Grosso, N., McCaughrean, M., Broos, P., Garmire, G.,
Townsley, L.\ 2005, \apjs, submitted
\bibitem[Glassgold et al.(2000)]{Glassgold00} Glassgold, A.~E.,
Feigelson, E.~D., \& Montmerle, T.\ 2000, in Protostars and
Planets IV (V.\ Mannings et al., eds.), Univ.\ Arizona Press, 429
\bibitem[Grosso et al.(2005)]{Grosso05} Grosso, N., Feigelson, E.\ D.,
and 13 others 2005, \apjs, submitted
\bibitem[G{\" u}del(2004)]{Guedel04} G{\" u}del, M.\ 2004,
\aapr, 12, 71
\bibitem[Helfand \& Moran(2001)]{Helfand01} Helfand, D.~J., \& 
Moran, E.~C.\ 2001, \apj, 554, 27
\bibitem[Hillenbrand(1997)]{Hillenbrand97} Hillenbrand, L.~A.\ 1997,
\aj, 113, 1733
\bibitem[Hillenbrand \& Hartmann(1998)]{Hillenbrand98} Hillenbrand,
L.~A., \& Hartmann, L.~W.\ 1998, \apj, 492, 540
\bibitem[Hong et al.(2004)]{Hong04} Hong, J., Schlegel, E.~M., 
\& Grindlay, J.~E.\ 2004, \apj, 614, 508 
\bibitem[Johnstone \& Bally(1999)]{Johnstone99} Johnstone, D., \&
Bally, J.\ 1999, \apjl, 510, L49
\bibitem[Lada et al.(2004)]{Lada04} Lada, C.~J., Muench, 
A.~A., Lada, E.~A., \& Alves, J.~F.\ 2004, \aj, 128, 1254
\bibitem[Lynden-Bell(1967)]{LyndenBell67} Lynden-Bell, D.\ 1967,
\mnras, 136, 101
\bibitem[Montmerle(1987)]{Montmerle87} Montmerle, T.\ 1987, in 
Starbursts and Galaxy Evolution (Thuan, T.\ X.\ et al., eds), 
Editions Frontieres, 47
\bibitem[Montmerle \& Grosso(2002)]{Montmerle02} Montmerle, T., \& 
Grosso, N.\ 2002, in The Origins of Stars and Planets: The VLT View 
(J.\ F.\ Alves \& M.\ J.\ McCaughrean, eds.), Eur Southern Obs, 453 
\bibitem[Muno et al.(2004)]{Muno04} Muno, M.~P., et al.\ 2004,
\apj, 613, 326
\bibitem[O'Dell(2001)]{ODell01} O'Dell, C.~R.\ 2001, \araa, 39,
99
\bibitem[Preibisch(2003)]{Preibisch03} Preibisch, T.\ 2003, \aap, 
410, 951 
\bibitem[Preibisch et al.(2005a)]{Preibisch05a} Preibisch, T., Kim,
Y.-C., Favata, F., and 8 others 2005, \apjs, submitted
\bibitem[Preibisch et al.(2005b)]{Preibisch05c} Preibisch, T.,
McCaughrean, M., and 7 others 2005, \apjs, submitted
\bibitem[Preibisch \& Feigelson(2005)]{Preibisch05b} Preibisch, T.
\& Feigelson,, E.\ D.\ 2005, \apjs, submitted
\bibitem[Ryter(1996)]{Ryter96} Ryter, C.~E.\ 1996, \apss, 236,
285
\bibitem[Scally \& Clarke(2002)]{Scally02} Scally, A., \& 
Clarke, C.\ 2002, \mnras, 334, 156 
\bibitem[Stelzer et al.(2005)]{Stelzer05} Stelzer, B., Flaccomio, E.,
Montmerle, T., Micela, G., Sciortino, S., Favata, F., Preibisch,
T., Feigelson, E.\ D.\ 2005, \apjs, submitted
\bibitem[Stevens et al.(1992)]{Stevens92} Stevens, I.~R., 
Blondin, J.~M., \& Pollock, A.~M.~T.\ 1992, \apj, 386, 265 
\bibitem[Townsley et al.(2003)]{Townsley03} Townsley, L.~K.,
Feigelson, E.~D., Montmerle, T., Broos, P.~S., Chu, Y., \&
Garmire, G.~P.\ 2003, \apj, 593, 874
\bibitem[Vuong et al.(2003)]{Vuong03} Vuong, M.~H., Montmerle,
T., Grosso, N., Feigelson, E.~D., Verstraete, L., \& Ozawa, H.\
2003, \aap, 408, 581
\bibitem[van der Werf \& Goss(1989)]{vanderWerf89} van der Werf,
P.~P., \& Goss, W.~M.\ 1989, \aap, 224, 209
\bibitem[Wilms et al.(2000)]{Wilms00} Wilms, J., Allen, A., \& 
McCray, R.\ 2000, \apj, 542, 914 
\end{thebibliography}
\end{document}